\documentclass[a4paper, 12pt]{article}
\usepackage [a4paper,left=2.5cm,bottom=2.5cm,right=2.5cm,top=2.5cm]{geometry}
\usepackage[english]{babel}
\usepackage{amssymb}
\usepackage{amsmath, amsthm, dsfont}
\usepackage{subcaption}
\usepackage{tabularx,array}
\usepackage{comment}
\usepackage{mathtools}  
\usepackage{graphicx, url}
\usepackage{enumitem} 
\usepackage{bm}
\usepackage{float}
\usepackage{bbm}
\usepackage{xcolor}
\usepackage{graphicx}
\usepackage{todonotes}
\usepackage{todonotes}
\usepackage{hyperref}
\usepackage{xfrac}
\hypersetup{colorlinks=true, linkcolor=blue,citecolor=gray}
\usepackage[nameinlink,capitalise]{cleveref}
\crefformat{equation}{#2(#1)#3}

\usepackage{color,soul}

\theoremstyle{plain}
\theoremstyle{plain}\newtheorem{assumption}{Assumption}
\crefname{assumption}{Assumption}{Assumptions}
\newtheorem{theorem}{Theorem}[section]
\newtheorem{lemma}[theorem]{Lemma}
\newtheorem{corollary}[theorem]{Corollary}
\newtheorem{proposition}[theorem]{Proposition}

\newtheorem{remark}[theorem]{Remark}
\newtheorem{definition}{Definition}

\newcommand{\E}{\mathbb{E}}

\newcommand{\R}{\mathbb{R}}

\renewcommand{\P}{\mathbb{P}}

\usepackage{autonum}
\numberwithin{equation}{section}

\title{On goodness-of-fit testing for volatility in McKean–Vlasov models}

\author{Akram Heidari \thanks{Department of Mathematics, University of Luxembourg, Maison du Nombre, 6 Avenue de la Fonte, 4364 Esch-sur-Alzette,
Luxembourg, Email: akram.heidari@uni.lu} \and Mark Podolskij \thanks{Department of Mathematics, University of Luxembourg, Maison du Nombre, 6 Avenue de la Fonte, 4364 Esch-sur-Alzette,
Luxembourg, Email: mark.podolskij@uni.lu}}

\date{}

\begin{document}
\maketitle

\begin{abstract}

This paper develops a statistical framework for goodness-of-fit testing of volatility functions in i.i.d. McKean–Vlasov stochastic differential equations, which model large  diffusion systems with distribution-dependent dynamics. Although integrated volatility estimation for classical SDEs is well established, formal model validation and goodness-of-fit testing for McKean–Vlasov systems remain largely unexplored, particularly in settings combining large particle limits with high-frequency observations.
We propose a goodness-of-fit test based on discrete observations of a particle system and study its asymptotic properties in a joint regime where both the number of particles and the sampling frequency tend to infinity. We establish asymptotic normality for the relevant volatility estimators and derive a central limit theorem for the resulting test statistic. These results provide a rigorous basis for assessing volatility specifications in high-dimensional mean-field diffusion models. \\

\noindent
 \textit{Keywords:}  Asymptotic distribution, goodness-of-fit testing, high-frequency data, McKean–Vlasov diffusions, mean-field models, nonlinear diffusion, volatility specification. 
 \\ 
 
\noindent
\textit{AMS subject classifications:} 
62E20, 
62G10,      
60F15,  	
60H10.  	
\\

\end{abstract}

\tableofcontents

\section{Introduction} \label{introduction}

Independent and identically distributed McKean--Vlasov stochastic differential equations have emerged as a fundamental mean-field framework for modelling complex systems composed of large populations of interacting agents. Unlike classical diffusion models, the dynamics of each particle depend not only on its current state but also on its probability distribution. This distributional dependence enables McKean--Vlasov models to capture collective and emergent phenomena arising from interactions within large systems, making them particularly suitable for applications in economics, finance, physics, and engineering \cite{Carmona1, DGZZ, Jourdain, Szn91}.

In financial applications, McKean--Vlasov dynamics have been employed to study systemic risk \cite{Fouque}, mean-field interactions in portfolio optimisation \cite{Carmona2}, and the evolution of agent-based financial markets. More generally, they play a central role in mean-field games, optimal control, equilibrium analysis, and the modelling mean-field limits of large-scale interacting systems; see, for instance, \cite{CarmonaWebster, Lacker}. A key advantage of these models is their ability to represent endogenous feedback mechanisms, whereby individual decisions and aggregate system behaviour are coupled through the evolving distribution of the population.

The increasing use of McKean--Vlasov models in applications has created a corresponding demand for statistical tools capable of validating their structural assumptions. In particular, the volatility function is of central importance, as it determines the stochastic fluctuations of the system and directly affects uncertainty quantification, prediction, and risk assessment. While volatility specifications in classical diffusion models typically depend only on state variables or time, such assumptions are often inadequate in mean-field settings where interactions between agents influence the dynamics. In these situations, the volatility coefficient may depend explicitly on the population distribution, and misspecification can lead to substantial modelling errors.

Despite the importance of model validation, goodness-of-fit testing for McKean--Vlasov systems remains largely unexplored. Existing testing procedures for volatility functions have been developed primarily for classical diffusion models \cite{Ait, CW99, DP08, DPV06} and fractional SDEs \cite{PW13}, where the coefficients depend only on finite-dimensional state variables. On the other hand, recent research on McKean--Vlasov equations and interacting particle systems has focused predominantly on parametric and nonparametric estimation problems \cite{ABPPZ23,First-paper,BPP22,B11,C21,CG23,CGL24,Marc,GL21a,GL21b,K90,L22,NPR24,SKPP21,WWMX16}, with particular emphasis on drift estimation. To the best of our knowledge, no general statistical methodology currently exists for assessing the adequacy of volatility specifications in i.i.d. McKean--Vlasov systems based on discrete observations.

The objective of this paper is to fill this gap by developing a goodness-of-fit testing framework for volatility functions in i.i.d. McKean--Vlasov models. We consider a system of $N$ i.i.d. particles $(X^i)^{i=1, \ldots, N}$, defined on a filtered probability space $(\Omega, \mathcal{F},(\mathcal{F}_t)_{t\geq 0},\P )$ satisfying the usual conditions, and evolving over a fixed time interval $[0, T]$. The particles are modelled as independent copies of a nonlinear process satisfying the McKean--Vlasov SDE
\begin{equation} \label{model fit-test}
\begin{cases}
d X_t^{i} = b \big(X_t^{i}, \mu_t \big) dt + a\big(X_t^{i}, \mu_t \big) d W_t^i  \qquad i = 1, \ldots , N, \quad t\in [0, T] \\[1.5 ex]
\text{Law} \big( X_0^1, \ldots , X_0^{N} \big) : = \mu_0 \times \ldots \times \mu_0
\end{cases}
\end{equation}
where the processes $(W^i_t)_{t \in [0, T]}$, $i=1,\ldots,N$, are independent Brownian motions, which are independent of the law of $ (X_0^1, \ldots , X_0^{N})$, and $\mu_t$ denotes the law of $X_t^i$. The model coefficients
$$
b: \R \times \mathcal{P}_2 \rightarrow \R, \qquad a: \R \times \mathcal{P}_2 \rightarrow \R
$$
are measurable functions depending on the current state and the current distribution of the solution. Here, $\mathcal{P}_2$ denotes the space of probability measures on $\R$ with finite second moments. This space is equipped with the Wasserstein-2 metric $W_2(\mu, \nu)$, where 
\begin{equation}\label{wass}
W_p(\mu, \nu) = \Big( \inf_{m \in \Gamma (\mu, \nu)} \int_{\R^2} |x - y|^p m(dx, dy) \Big)^{\frac 1 p},
\qquad p\geq 1,
\end{equation}
and $\Gamma(\mu, \nu)$ denotes the set of probability measures on $\R^2$ with marginals $\mu$ and $\nu$.

Our statistical analysis is based on discrete observations of the particle system,
\begin{equation} \label{data}
\left(X_{t_{j}}^{i}\right)_{j = 1, \ldots, n}^{i = 1, \ldots, N}, \qquad \text{with }   t_{j} = T j / n,
\end{equation}
and is carried out in a joint asymptotic regime where both the observation frequency increases, $\Delta_n := T/n \to 0$, and the number of particles diverges, $N\to\infty$, while the time horizon $T>0$ remains fixed.

Our primary goal is to test the null hypothesis that the volatility function $a(x,\mu)$ belongs to a given parametric family. More precisely, we aim at testing whether $a^2(x,\mu_t) = \sum_{k=1}^d \lambda_k a^2_{k}(x,\mu_t)$, for given functions $a_1^2,\ldots, a_d^2$ and some $\lambda_k$'s, $\mu_t(dx)dt$--almost surely.  
To this end, we construct a goodness-of-fit statistic based on a suitable distance measure between the unobserved volatility and the vector space generated by $a_1^2,\ldots, a_d^2$. We establish asymptotic normality of the underlying volatility estimators under the additional condition $N\Delta_n^2\to 0$ and derive a central limit theorem for the proposed test statistic. These results yield an asymptotically valid testing procedure that attains the prescribed significance level under the null hypothesis and is consistent against fixed alternatives.

The principal methodological challenge stems from the distributional dependence of the volatility coefficient. This feature induces nonlinear interactions and path-dependent effects that are absent in classical diffusion models and prevents a direct application of standard techniques from high-frequency statistics. Our analysis combines high-frequency asymptotics with mean-field arguments to overcome these difficulties and establish the asymptotic behaviour of the proposed procedure.

To the best of our knowledge, this paper provides the first rigorous goodness-of-fit testing framework for volatility functions in i.i.d. McKean--Vlasov models based on discrete observations of interacting particle systems. The results extend the scope of statistical model validation to diffusion models with distribution-dependent dynamics and provide a foundation for hypothesis testing in nonlinear mean-field systems.

The remainder of the paper is organised as follows. Section \ref{assumption} introduces the modelling framework and the standing assumptions. Section \ref{test-vol} develops the proposed testing methodology and presents the construction of the test statistic. Section \ref{main-result} contains the main theoretical results, including asymptotic normality of the estimators and the limiting distribution of the test statistic under the null hypothesis. All proofs are deferred to Section \ref{proofs}.

\subsubsection*{Notation}
Throughout, $C$ denotes a generic positive constant whose value may change from line to line. When its dependence on a parameter $p$ is relevant, we write $C_p$. The notation $O_{\mathbb P}(1)$ (respectively, $o_{\mathbb P}(1)$) denotes a sequence of random variables that is stochastically bounded (respectively, converges to zero in probability) as $N\to\infty$ and $\Delta_n\to0$. More generally, for a deterministic sequence $r_{n,N}>0$ satisfying either $r_{n,N}\to0$ or $r_{n,N}\to\infty$, the notations $O_{\mathbb P}(r_{n,N})$ and $o_{\mathbb P}(r_{n,N})$ are defined similarly. For $A\in\mathbb R^{d\times d}$, $\mathrm{vec}(A)\in\mathbb R^{d^2}$ denotes the vectorization of $A$, obtained by stacking its columns into a single vector. For $x\in\mathbb R^d$, we write $\|x\|_{\max}$ for the maximum norm of $x$ and $x^{\otimes m}$ for its $m$-fold tensor product. For a multivariate function $f$, we denote its partial derivatives by
$\partial_x f$, $\partial_{xy} f$, and so on.

\section{Assumptions}\label{assumption}
In this section, we introduce the main assumptions associated with the model \eqref{model fit-test}.
\begin{assumption} \label{assumpt1}
The initial distribution $\mu_0$ possesses the Lebesgue density $p_0$, which satisfies the lower bound
\[
p_0(x) \geq c_0\exp(-c_1 x^2), \qquad x\in \R,
\]
for some constants $c_0,c_1>0$. Furthermore,
for all $k \ge 1$,
$$ \int_{\R} |x|^k \mu_0 (d x) \le C_k. $$
\end{assumption}

The following assumption ensures the existence and uniqueness of a strong solution to \eqref{model fit-test}, guaranteeing well-posedness of the model. Along with Assumption \ref{assumpt1} it also guarantees existence of moments of all orders.

\begin{assumption}\label{assumpt2}
The drift and diffusion coefficients satisfy Lipschitz continuity and a linear growth condition. Specifically, there exists a constant $C > 0$ such that for all $(x,\mu), (y,\lambda) \in \R \times {\mathcal{P}_2}$:
\begin{align}
& |b(x,\mu) - b(y,\lambda)| + |a(x,\mu) - a(y,\lambda)| \leq C \big( |x - y| + W_2(\mu,\lambda) \big), \\
& |b(x,\mu)|^2 + |a(x,\mu)|^2 \leq C \big( 1 + |x|^2 + W_2^2(\mu, \delta_0) \big).
\end{align}
Furthermore, the functions $a(\cdot,\mu), b(\cdot,\mu)\in C^2(\R)$ satisfy 
$\|\partial_{xx}a(x,\mu)\|_{\infty}, \|\partial_{xx}b(x,\mu)\|_{\infty} \leq C$ for all $\mu \in \mathcal{P}_2$
and it holds that 
\begin{align}
|a(x,\mu)| \geq v>0, \qquad \forall x\in \R,~ \mu \in \mathcal{P}_2.
\end{align}
\end{assumption}


\noindent
Regarding the regularity of the diffusion function $ a: \R \times \mathcal{P}_2 \rightarrow \R $, we adopt the notion of linear differentiability, which is widely used in the literature on McKean–Vlasov equations and mean-field games to characterize the smoothness of the mapping $ \mu \mapsto a (x, \mu)$ from
$ \mathcal{P}_2 \rightarrow \R $. This concept is particularly well-suited to our framework, and we refer the reader to Section 2 of \cite{Marc} and the references therein for a detailed exposition.
\begin{definition} \label{derivat of mu}
A mapping $ f:\R\times \mathcal{P}_2 \rightarrow \R $ is said to have a linear functional derivative, if there exists $ \partial_{\mu} f: \R^2 \times \mathcal{P}_2 \rightarrow \R $ such that
\begin{equation}
f(x,\mu)-f(x,\mu^\prime) = \int_0^1 \int_{\R} \partial_\mu f(x,y, \lambda \mu + (1-\lambda) \mu^\prime) (\mu - \mu^\prime) (d y) d \lambda
\end{equation}
for every $x\in \R$, $(\mu, \mu^\prime) \in \mathcal{P}_2 $ and $\partial_{\mu} f$ satisfies additional smoothness properties, which will be provided in the following assumption.
\end{definition}


\begin{assumption} \label{assumpt3}
The map $\mu \mapsto a( x, \mu)$  admits a functional derivative in the sense of Definition~\ref{derivat of mu}. Furthermore, there exists a constant $C>0$ such that for all $(x, \mu), (x^{\prime}, \mu^{\prime}) \in \R \times \mathcal{P}_2 $,
\begin{align}
& \big|\partial_{\mu} a( x, y, \mu)- \partial_{\mu} a( x^\prime, y^\prime, \mu^\prime) \big| \le C (|x-x^\prime| + |y-y^\prime| + W_2 (\mu,\mu^\prime)).
\end{align}
Additionally it holds that 
\begin{align}
    &|\partial_y\partial_{\mu} a( x, y, \mu)| +
    |\partial_{yy}\partial_{\mu} a( x, y, \mu)|\leq C \qquad \forall ( x, y, \mu), \\
    &|\partial_y\partial_{\mu} a( x, y, \mu) - \partial_y\partial_{\mu} a( x, y, \mu')|\leq C 
    W_1 (\mu,\mu^\prime) \qquad \forall ( x, y, \mu).
\end{align}
\end{assumption} 

We first note that Assumptions \ref{assumpt1} and \ref{assumpt2} imply that, for every
$t\in[0,T]$, the law $\mu_t$ of $X_t$ admits a density $p_t$ with respect to the
Lebesgue measure. Moreover, $(p_t)_{t\in[0,T]}$ solves the nonlinear Fokker--Planck
equation (cf. \cite{F05})
\[
\partial_t p_t(x)
=
-\partial_x\!\big(b(x,\mu_t)p_t(x)\big)
+\frac12\,\partial_{xx}\!\big(a^2(x,\mu_t)p_t(x)\big).
\]
Furthermore, due to Assumption \ref{assumpt3} the map
\[
a(x,t):=a(x,\mu_t)
\]
belongs to $C^{2,1}(\mathbb{R}\times[0,T])$. Indeed, applying the chain rule on the
Wasserstein space yields
\[
\partial_t a(x,t)
=
\int_{\mathbb R}
\partial_\mu a(x,y,\mu_t)\,
\partial_t p_t(y)\,dy.
\]
Substituting the Fokker--Planck equation and integrating by parts, and using Assumption \ref{assumpt3}, one concludes that the above integral is finite.

As a consequence, the process $(a(X_t^i,t))_{t\in[0,T]}$ is a continuous
semimartingale. Applying It\^o's formula gives
\begin{align}
a(X_t^i,t)
&=
a(X_0^i,0)
+
\int_0^t
\Big(
\partial_t a(X_s^i,s)
+
b(X_s^i,\mu_s)\,\partial_x a(X_s^i,s)
+
\frac12\,a^2(X_s^i,s)\,\partial_{xx}a(X_s^i,s)
\Big)\,ds
\nonumber\\
\label{Itoassa}
&\quad
+
\int_0^t
a(X_s^i,s)\,\partial_x a(X_s^i,s)\,dW_s^i .
\end{align}
This semimartingale representation plays a crucial role in the derivation of error
estimates for high-frequency statistics associated with the process
$(X_t^i)_{t\in[0,T]}$; see, for instance, \cite{BGJPS}. In relation to functions appearing in 
\eqref{Itoassa},  Assumptions~\ref{assumpt1}--\ref{assumpt3}, guarantees that they exhibit at most polynomial growth in the variable $x$. Since the process $(X_t^i)_{t\in[0,T]}$ admits moments of all orders, it follows that the drift and diffusion coefficients of the process $(a(X_t^i,t))_{t\in[0,T]}$ also possess finite moments of any order.

\begin{remark} \label{Examp} \rm
A common class of measure-dependent volatility functions considered in the literature is given by
\[
a(x,\mu)
=
f\!\left(x,\int_{\mathbb{R}}\psi(z)\,\mu(dz)\right),
\]
where \(f:\mathbb{R}^2\to\mathbb{R}\) and \(\psi:\mathbb{R}\to\mathbb{R}\) are sufficiently smooth functions. Such examples are common in the theoretical works 
as in e.g. \cite{F84,M66}. 
In this case, a direct application of the chain rule for 
the linear functional derivative yields
\[
\partial_\mu a(x,y,\mu)
=
\partial_{x_2}f\!\left(x,\int_{\mathbb{R}}\psi(z)\,\mu(dz)\right)\psi(y),
\]
where \(\partial_{x_2}f\) denotes the partial derivative of \(f\) with respect to its second argument. Consequently, Assumptions~\ref{assumpt1}--\ref{assumpt3} can be verified directly under suitable regularity and growth conditions on \(f\) and \(\psi\). We remark however that there exist models in the literature, where the function $\psi$ is not smooth. We refer e.g. to \cite{HL17}, whose model corresponds to $\psi=1_{(-\infty,0]}$.   
\qed
\end{remark}

\section{ Testing parametric hypotheses for the volatility}\label{test-vol}

In this section, we develop a goodness-of-fit testing framework for the volatility structure in McKean–Vlasov SDEs. Our goal is to assess whether a given parametric form of the volatility function is consistent with the observed behavior of a discretely sampled particle system. We begin by formally stating the parametric hypothesis. Then, we introduce our proposed test statistic $\widehat{S}^N$ and describe its construction under high-frequency and large-population asymptotics.


Let  
\[
a_1, \ldots, a_d : \R \times \mathcal{P}_2 \to \R
\]  
be a collection of known functions, assumed to be linearly independent and to satisfy the same regularity conditions as the volatility function \(a(x,\mu)\). Loosely speaking, our objective is to test whether the squared volatility function 
\(a^2\) belongs to the linear span of the basis functions 
\(a_1^2, \ldots, a_d^2\) (a more precise formulation is presented in Remark \ref{Rem3.1}). More specifically, the null hypothesis is given by  
\begin{equation} \label{test}
H_0 : L := \min_{(\lambda_1, \ldots, \lambda_d) \in \R^d} 
\int_0^T \int_{\R} 
\Big( a^2(x, \mu_t) - \sum_{k=1}^d \lambda_k a^2_k(x, \mu_t ) \Big)^2 
\, \mu_t(dx)\, dt = 0,
\end{equation}  
with the alternative hypothesis \(H_1 : L > 0\).  
Here, \(\mu_t\) denotes the distribution of the underlying particles 
\(X_t^1, \ldots, X_t^N\), which is not directly observable. In practice, we approximate \(\mu_t\) by the empirical distribution of the observed particles. The criterion in \eqref{test} serves as a natural foundation for our test, since it directly measures model discrepancy in a Hilbert space framework. Moreover, it admits a discretized version that can be readily implemented using particle observations.

\begin{remark} \label{Rem3.1} \rm
The distance measure $L$ introduced in \eqref{test} is conceptually related to the distance proposed in \cite{DP08,DPV06} for classical SDEs, although the two approaches differ in essential aspects. To clarify this, recall that \cite{DP08,DPV06} study the one-dimensional diffusion model  
\[
dX_t = b(X_t)\,dt + a(X_t)\,dW_t ,
\]
observed at discrete time points $t_{j}$. They introduce the \textit{random} 
distance measure  
\[
M^2 := \min_{(\lambda_1, \ldots, \lambda_d) \in \R^d} 
\int_0^T \Big( a^2(X_t) - \sum_{k=1}^d \lambda_k a_k^2(X_t) \Big)^2 \, dt ,
\]
and consider the hypothesis test $H_0: M^2 = 0$ versus $H_1: M^2 > 0$. 
A key point is that, under high-frequency observations of a single trajectory of $(X_t)_{t \in [0,T]}$, one can only verify whether  
\[
a^2(X_t) = \sum_{k=1}^d \lambda_k a_k^2(X_t) 
\]
holds for some choice of $\lambda_k$'s, \emph{along the realized path} 
$(X_t(\omega))_{t \in [0,T]}$. There is no possibility to test this identity outside the observed trajectory, i.e., for 
$x \notin (X_t(\omega))_{t \in [0,T]}$.  

In contrast, in our setting with $N$ independent trajectories as in \eqref{model fit-test}, the condition $L = 0$ entails that  
\[
a^2(x,\mu_t) = \sum_{k=1}^d \lambda_k a_k^2(x,\mu_t)
\]
for some $\lambda_k$'s, holding for $\mu_t$-almost every $x \in \R$, $t\in [0,T]$. 
Nevertheless, this identity is testable only with respect to the distributions $\mu_t$ of the observed particles $(X_t^i)$, and not for arbitrary distributions.  
    \qed
\end{remark}

\noindent
Standard arguments (cf. \cite{Achi}) show that this $L^2$-distance admits the closed-form expression:
\begin{equation}\label{stat1}
L= g(\Gamma_1, \ldots, \Gamma_d,  \mathcal{B}, \Lambda)= \mathcal{B} - (\Gamma_1, \ldots, \Gamma_d) \Lambda^{-1} (\Gamma_1, \ldots, \Gamma_d)^{\top}
\end{equation}
where the quantities $\mathcal{B}, \Gamma_1, \ldots, \Gamma_d$ and the matrix $\Lambda = (\Lambda_{k,l})_{1 \leq k,l \leq d}$ are given by
\begin{align} \label{stat2}
\begin{split}
& \mathcal{B}:= \int_0^T \int_{\R} a^4(x, \mu_t) \mu_t(dx)  dt, \\
& \Gamma_k:= \int_0^T \int_{\R} a_k^2(x, \mu_t) a^2(x, \mu_t) \mu_t(dx)  dt , \qquad k=1, \ldots, d \\
& \Lambda_{k,l}:= \int_0^T \int_{\R} a_k^2(x, \mu_t) a_{l}^2(x, \mu_t) \mu_t(dx)  dt , \qquad k,l=1, \ldots, d 
\end{split}
\end{align}

\begin{remark} \rm
(Invertability of $\Lambda$)
Under Assumptions \ref{assumpt1} and \ref{assumpt2}, the transition density
$p(s,x;t,y)$, $0\leq s<t\leq T$, satisfies Aronson-type estimates. More precisely,
there exist constants $q_1,\ldots,q_4>0$ such that
\[
\frac{q_1}{\sqrt{t-s}}
\exp\!\left(
-q_2\frac{(x-y)^2}{t-s}
\right)
\leq
p(s,x;t,y)
\leq
\frac{q_3}{\sqrt{t-s}}
\exp\!\left(
-q_4\frac{(x-y)^2}{t-s}
\right).
\]
(cf \cite{A67}).
In particular, the lower bound implies that
\[
p(s,x;t,y)>0,
\qquad 0\leq s<t\leq T,\quad x,y\in\mathbb R.
\]
Consequently, since the initial density $p_0$ fulfills the same assumption, the marginal density
$p_t$ satisfies
\[
p_t(x)>0,
\qquad x\in\mathbb R,\quad t\in[0,T].
\]
Let now $v=(v_1,\ldots,v_d)^\top\in\mathbb R^d$. Then
\[
v^\top \Lambda v
=
\int_0^T\int_{\mathbb R}
\left(
\sum_{k=1}^d v_k a_k^2(x,\mu_t)
\right)^2
p_t(x)\,dx\,dt.
\]
Since $p_t(x)>0$ for all $(t,x)\in(0,T]\times\mathbb R$, we have
$v^\top \Lambda v=0$ if and only if
\[
\sum_{k=1}^d v_k a_k^2(x,\mu_t)=0
\]
for almost all $(t,x)\in(0,T]\times\mathbb R$.
By the linear independence of the functions
$a_1^2,\ldots,a_d^2$, this is only possible when
$v=0$. Therefore,
\[
v^\top \Lambda v>0,
\qquad v\neq 0,
\]
showing that $\Lambda$ is positive definite and, in particular, invertible.
\qed
\end{remark}

\noindent
In order to construct a consistent estimator of $L$ in \eqref{stat1}, we replace the quantities in \eqref{stat2} with their empirical counterparts, based on the discrete-time observations introduced in \eqref{data}. Accordingly, we define the following estimators:
\begin{align} \label{estimators}
\begin{split}
& \widehat{\mathcal{B}}:= \frac{1}{3 N \Delta_n} \sum_{i=1}^N \sum_{j=1}^n |X_{t_{j+1}}^i - X_{t_j}^i |^4 \\
& \widehat{\Gamma}_k := \frac{1}{N} \sum_{i=1}^N \sum_{j=1}^n a_k^2(X_{t_j}^i, \mu_{t_j}^N) |X_{t_{j+1}}^i - X_{t_j}^i |^2  \qquad \quad k=1, \ldots, d\\
& \widehat{\Lambda}_{k,l}:= \frac{\Delta_n}{N} \sum_{i=1}^N \sum_{j=1}^n a_k^2(X_{t_j}^i, \mu_{t_j}^N) a_l^2(X_{t_j}^i, \mu_{t_j}^N) \qquad \quad k,l=1, \ldots, d
\end{split}
\end{align}
(For simplicity of notation we suppress the dependence of estimators on $n$ and $N$). Here, $\mu_t^N$ denotes the empirical measure of the system at time $t$, i.e.\
\[ \mu_t^N := \frac{1}{N} \sum_{i = 1}^N \delta_{X_t^i}. \]
We then introduce the following test statistic as the empirical analogue of the quantity \eqref{stat1}:
\begin{equation}
\widehat{S}^N:= g(\widehat{\Gamma}_1, \ldots, \widehat{\Gamma}_d, \widehat{\mathcal{B}}, \widehat{\Lambda})= \widehat{\mathcal{B}} - \widehat{\mathbf{\Gamma}}^\top \widehat{\Lambda}^{-1} \widehat{\mathbf{\Gamma}} 
\end{equation}
where $\widehat{\mathbf{\Gamma}}= (\widehat{\Gamma}_1, \ldots, \widehat{\Gamma}_d)^\top $ and $\widehat{\Lambda}=(\widehat{\Lambda}_{k,l})_{1\leq k,l\leq d}$ is defined by the components in \eqref{estimators}. This statistic captures the deviation from the null hypothesis and forms the basis of our goodness-of-fit test.

In the following section, we  summarise the key statistical properties of our proposed estimator $\widehat{S}^N$. Specifically, building on the consistent estimation of $ \mathcal{B} $, $ \Gamma_k $, and $ \Lambda_{k,l} $ by $ \widehat{\mathcal{B}} $, $ \widehat{\Gamma}_k $, and $ \widehat{\Lambda}_{k,l}$, we show a central limit theorem for the estimator $ \widehat{S}^N $.

\section{Main results}\label{main-result}

In this section, we present the core theoretical contributions of this paper, focusing on the asymptotic properties of our proposed estimators and their associated test statistics. 
We begin by demonstrating the consistency and convergence rates of our estimators. Specifically, Theorem~\ref{thrm1} establishes the consistent approximation of the limiting matrix $\Lambda$ by its empirical counterpart $\widehat{\Lambda}$. Similarly, Theorem~\ref{thrm2} provides stochastic expansions for the quantities $\widehat{\mathcal{B}}$ and $\widehat{\Gamma}_k$ (for $k=1, \ldots, d$), which are essential for deriving the limiting distribution of our test statistic. Corollary~\ref{thrm3} then establishes the joint asymptotic normality of these key components. These results together characterize the limiting distribution and its asymptotic covariance structure. Finally, combining these foundational results, we present the main asymptotic normality result for our proposed estimator $\widehat{S}^N$. 

In the following, we establish stochastic expansions for the estimators
\(\widehat{\Lambda}\), \(\widehat{\mathcal B}\), and \(\widehat{\Gamma}_k\).
Before stating the main results, we introduce some additional notation.
For sufficiently smooth functions \(f,g:\mathbb R\times\mathcal P_2\to\mathbb R\), define
\begin{align}\label{defih}
h_t(f,g)(x)
:={}&
fg(x,\mu_t)
+\E\!\left[\partial_\mu(fg)(X_t^1,x,\mu_t)\right]  \\
&\quad
-\int_{\mathbb R^2}
\partial_\mu(fg)(z,y,\mu_t)\,
\mu_t(dz)\mu_t(dy).
\nonumber
\end{align}
We are now in a position to state the main result of the paper.

\begin{theorem}\label{thrm1}
Assume that Assumptions \ref{assumpt1}-\ref{assumpt3} hold and $N\Delta_n^2\to 0$. Then
\begin{equation}
\sqrt{N} \big(\widehat{\Lambda}- \Lambda \big) =\sqrt{N} M_{\Lambda} + o_{\P}(1)
\end{equation}
where $M_\Lambda$ is a $(d \times d)$-matrix with elements 
\begin{align}
& M_{\Lambda,k,l}:= \frac{1}{N} \sum_{i=1}^N \big( Z_{\Lambda,k,l}^i - \E(Z_{\Lambda,k,l}^1) \big) \\
& Z_{\Lambda,k,l}^i:= \int_0^T h_s(a^2_k,a^2_l) (X_s^i)d s \qquad \qquad  k,l = 1, \ldots, d,
\end{align}
where the function $h_t$ is defined at  \eqref{defih}.
\end{theorem}

\begin{theorem}\label{thrm2}
Assume that Assumptions \ref{assumpt1}-\ref{assumpt3} hold and $N\Delta_n^2\to 0$. Then
\begin{align}
\sqrt{N} \Big( \widehat{\Gamma}_k - \Gamma_k \Big) = \sqrt{N} M_k + o_{\P}(1)
\label{thrm2.1}\\
\sqrt{N} \Big( \widehat{\mathcal{B}} - \mathcal{B} \Big) = \sqrt{N} M_{\mathcal{B}} + o_{\P}(1)
\label{thrm2.2}
\end{align}
with
\begin{align}
& M_k:= \frac{1}{N} \sum_{i=1}^N \big( Z_k^i - \E(Z_k^1) \big), \qquad Z_k^i:= \int_0^T 
a^2(X_s^i,\mu_s) h_s(a^2_k,1)(X_s^i)d s, \\
& M_{\mathcal{B}}:= \frac{1}{N} \sum_{i=1}^N \big( Z_{\mathcal{B}}^i - \E(Z_{\mathcal{B}}^1) \big), \qquad Z_{\mathcal{B}}^i:= \int_0^T a^4(X_s^i,\mu_s) d s,
\end{align}
where the function $h_t$ is defined at  \eqref{defih}.
\end{theorem}

\noindent
We note that the terms $Z_{\Lambda,k,l}^i$ differ substantially in structure from $Z_k^i$ and $Z_{\mathcal{B}}^i$. The reason is that the statistics $\widehat{\Gamma}_k$ and $\widehat{\mathcal{B}}$ are based on the increments $X_{t_{j+1}}^i-X_{t_j}^i$, which admit the approximation
\[
X_{t_{j+1}}^i-X_{t_j}^i
\approx
a(X_{t_j}^i,\mu_{t_j})\bigl(W_{t_{j+1}}^i-W_{t_j}^i\bigr).
\]
Hence, the latter depends on the true measure $\mu_{t_j}$ only. In contrast, the statistic $\widehat{\Lambda}$ explicitly involves the empirical measure $\mu_{t_j}^N$. The additional randomness introduced by replacing $\mu_{t_j}$ with $\mu_{t_j}^N$ gives rise to an extra asymptotically normal term, as is reflected in the definition \eqref{defih}.

Building upon the individual asymptotic properties established in Theorems \ref{thrm1} and \ref{thrm2}, we are now in a position to derive the joint asymptotic distribution of the involved components  $ (M_1, \ldots, M_d, M_{\mathcal{B}}, M_\Lambda) $. The following statement is a simple consequence of the standard central limit theorem and the $\delta$-method.

\begin{corollary} \label{thrm3}
Assume that  Assumptions \ref{assumpt1}-\ref{assumpt3} are satisfied and $N\Delta_n^2\to 0$.
\begin{itemize}
\item[(i)] It holds that 
\begin{equation} 
\widehat{Z}:=\sqrt{N} \left(\begin{array}{c} M_1 \\ \vdots \\ M_d \\ M_{\mathcal{B}} \\ \mathrm{vec}\;  (M_\Lambda)_{k,l} \end{array} \right) \xrightarrow{\mathcal{L}} \mathbf{Z}^* , \qquad \mathbf{Z}^* \buildrel d \over= \mathcal{N}_{d^2+d+1} (0, \Sigma)
\end{equation}
where the components of the covariance matrix $\Sigma$ are given by
\begin{equation} \label{matrix sigma}
\Sigma_{p,q}= \mathrm{Cov}\left(\widehat{Z}_p, \widehat{Z}_q\right).
\end{equation}
\item[(ii)] It holds that 
\begin{equation}
\sqrt{N} ( \widehat{S}^N - L ) \xrightarrow{\mathcal{L}} \mathcal{G} \sim \mathcal{N}(0,\tau^2)
\end{equation}
where the asymptotic variance $\tau^2$ is defined as
\[
\tau^2 = \nabla g^\top(\Gamma_1, \ldots, \Gamma_d, \mathcal{B}, \Lambda) \Sigma \nabla g(\Gamma_1, \ldots, \Gamma_d, \mathcal{B}, \Lambda).
\]
\end{itemize}
\end{corollary}

\noindent
The asymptotic result of Corollary~\ref{thrm3} forms the theoretical basis of our goodness-of-fit test. Under the null hypothesis \(H_0 : L = 0\), it yields  
\[
\sqrt{N}\, \widehat{S}^N \xrightarrow{\mathcal{L}} \mathcal{N}(0,\tau^2).
\]

\begin{remark} \rm
(Self-normalized statistic)
For practical applications, the null hypothesis will almost never hold exactly. 
It is therefore natural to ask how well the linear span of the functions 
\(a_1^2, \ldots, a_d^2\) can approximate the true squared volatility coefficient \(a^2\). 
The distance measure \(L\) is not ideal in this context, since its numerical size is 
difficult to interpret. A more convenient criterion was introduced in \cite{PW13} for 
fractional diffusion models and, in our setting, takes the form  
\[
G := \frac{L}{\mathcal{B}}.
\]  
In contrast to \(L\), the statistic \(G\) enjoys the appealing property 
\(G \in [0,1]\), which follows directly from Pythagoras’ theorem. 
This normalization allows deviations from the null hypothesis to be expressed in 
relative terms rather than in absolute units. Moreover, for any fixed 
\(\delta \in (0,1)\), one can test  
\[
H_0:~G \in [0,\delta] \qquad \text{vs.} \qquad H_1:~G \in (\delta,1],
\]  
and the asymptotic normality of \(G\) follows directly from Corollary~\ref{thrm3}.
 \qed
\end{remark}

\begin{remark} \rm
(Condition $N\Delta_n^2\to 0$) The condition \(N\Delta_n^2 \to 0\), which links the number of particles \(N\) to the sampling frequency \(\Delta_n\), is crucial for the validity of the main results of the paper. Indeed, if \(N\Delta_n^2 = O(1)\), a non-negligible asymptotic bias arises, for instance from the term \(H_{(1)}\) appearing in the proof of Theorem~\ref{thrm1}. This term reflects the error incurred by a Riemann-sum approximation and therefore does not vanish at the relevant asymptotic scale.

In principle, the exact asymptotic behavior of \(H_{(1)}\) can be derived using the results of \cite{BGJPS}. However, the resulting limit would depend on the drift and volatility coefficients in \eqref{Itoassa}. Since these quantities are typically unknown and difficult to estimate accurately from the available data, such a characterization would be of limited practical use. Similar consideration holds true for the case $N\Delta_n^2 \to \infty$. \qed
\end{remark}

\begin{remark} \rm
(Local alternatives)
Corollary~\ref{thrm3}(ii) provides a general weak convergence result that remains valid under both the null and alternative hypotheses. It can therefore be used to investigate the asymptotic power of the proposed test against local alternatives.

As an illustration, let $a_0,a_1,\ldots,a_d$ be linearly independent functions and suppose that the true volatility function is of the form
\[
a^2(x,\mu)
=
N^{-1/4}a_0^2(x,\mu)
+
\sum_{k=1}^d \lambda_k a_k^2(x,\mu),
\]
for some coefficients $\lambda_1,\ldots,\lambda_d$. In this setting, the null hypothesis is violated by the perturbation
$
N^{-1/4}a_0^2(x,\mu),
$
which corresponds to the critical rate since the resulting squared distance from the null model is of order $N^{-1/2}$.

Denote by $V$ the linear subspace generated by the functions
$a_1^2,\ldots,a_d^2$, and introduce the seminorm
\[
\|f\|_{\star}^2
:=
\int_0^T \int_{\mathbb R}
f^2(x,\mu_t)\,\mu_t(dx)\,dt.
\]
A straightforward calculation yields the identity
\[
L
=
N^{-1/2}
\min_{v\in V}
\|a_0^2-v\|_{\star}^2.
\]
Consequently, Corollary~\ref{thrm3}(ii) implies that
\[
\sqrt{N}\,\widehat S^N
\xrightarrow{\mathcal L}
\mathcal N\!\left(
\min_{v\in V}\|a_0^2-v\|_{\star}^2,
\,\tau^2
\right).
\]
Hence, the limiting distribution acquires a non-zero mean determined by the distance of $a_0^2$ from the null space $V$, which allows one to derive the asymptotic power of the test against local alternatives of this form.
\qed
\end{remark}

\subsection{Estimation of the asymptotic variance}
Due to the intricate structure of the test statistic, and in particular to its dependence on the empirical measure $\mu_t^N$, estimating the asymptotic covariance matrix $\Sigma$ introduced in Corollary~\ref{thrm3} is nontrivial. To address this issue, we propose a subsampling procedure.

Let $K=K_N$ be the number of  blocks, and partition the set of particle indices $\{1,\ldots,N\}$ into disjoint blocks
\[
B_1,\ldots,B_K,
\]
each containing $N/K$ particles. We assume that $K\to \infty$
and $N/K \to \infty$ as $N\to \infty$.
For each block $B_k$, $k=1,\ldots,K$, define
\begin{align} \label{defhatv}
\widehat{V}_k
:=
\left(
\widehat{\Gamma}_1^{(k)},
\ldots,
\widehat{\Gamma}_d^{(k)},
\widehat{\mathcal{B}}^{(k)},
\operatorname{vec}\bigl(\widehat{\Lambda}_{jl}^{(k)}\bigr)_{j,l}
\right)^\top.
\end{align}
Here, the estimators $\widehat{\Gamma}^{(k)},
\widehat{\mathcal{B}}^{(k)},
\widehat{\Lambda}^{(k)}$
are defined analogously to those in \eqref{estimators}, but are computed using only the particles $X^i$ with indices $i\in B_k$. In particular, all empirical measures appearing in these estimators are constructed solely from observations within the block $B_k$.
Since the blocks are disjoint and the particles are independent, the random vectors $\bigl(\widehat{V}_k\bigr)_{1\leq k\leq K}$
are independent and identically distributed. We further introduce the corresponding population vector
\begin{align} \label{defv}
V
:=
\left(
\Gamma_1,
\ldots,
\Gamma_d,
\mathcal{B},
\operatorname{vec}\bigl(\Lambda_{jl}\bigr)_{j,l}
\right)^\top,
\end{align}
and denote by $\widehat{V}$ the analogue of $\widehat{V}_k$ constructed from the full sample of $N$ particles.

We define the subsampling estimator of the asymptotic covariance matrix $\Sigma$ by
\begin{align}
\widehat{\Sigma}
:=
\frac{1}{K}
\sum_{k=1}^K
\frac{N}{K}
\bigl(\widehat{V}_k-\widehat{V}\bigr)
\bigl(\widehat{V}_k-\widehat{V}\bigr)^\top.
\label{eq:subsampling_covariance_estimator}
\end{align}
The main result of this subsection establishes the weak consistency of the estimator $\widehat{\Sigma}$.

\begin{proposition} \label{propSig}
Consider the subsampling statistic 
$\widehat{\Sigma}$ defined in \eqref{eq:subsampling_covariance_estimator}. Suppose that Assumptions \ref{assumpt1}-\ref{assumpt3} hold, $N\Delta_n^2\to 0$, $K\to \infty$ and $N/K\to \infty$. Then it holds
\[
\widehat{\Sigma} \xrightarrow{\P} \Sigma.
\]
\end{proposition}

\noindent
As a consequence of  Proposition \ref{propSig}, 
the estimator 
\[
\widehat{\tau}^2:= \nabla g^\top\left(\widehat{\Gamma}_1, \ldots, \widehat{\Gamma}_d, \widehat{\mathcal{B}}, \widehat{\Lambda}\right) \widehat{\Sigma} \nabla g\left(\widehat{\Gamma}_1, \ldots, \widehat{\Gamma}_d, \widehat{\mathcal{B}}, \widehat{\Lambda}\right)
\]
satisfies $\widehat{\tau}^2 \xrightarrow{\P} \tau^2$. Then, for a given significance level \(\alpha \in (0,1)\), the null hypothesis 
\(H_0 : L = 0\) is rejected whenever  
\[
\frac{\sqrt{N}\,\widehat{S}^N}{\widehat{\tau}} > z_{1-\alpha},
\]  
where \(z_{1-\alpha}\) denotes the \((1-\alpha)\)-quantile of the standard normal distribution. By construction, this test attains the correct asymptotic size \(\alpha\). Moreover, under the alternative \(H_1 : L > 0\), we have 
\(\sqrt{N}\,\widehat{S}^N \xrightarrow{\P} +\infty\), which ensures that the procedure is consistent against any fixed alternative.

\section{Proofs}\label{proofs}

As a preliminary step, we recall a collection of moment bounds, which are well known in the literature.

\begin{lemma} \label{l: momentsss}
Assumptions~\ref{assumpt1}–\ref{assumpt3} hold. Then, for any $p \ge 1$, there exists a constant $C_p>0$ such that the following bounds hold uniformly over all particles $i \in \{1, \dots, N\}$, for all $N \in \mathbb{N}$, and for all times $t \in [0, T]$: 
\begin{enumerate}
 \item[(i)] $ \displaystyle \sup_{t \in [0,T]} \mathbb{E}[|X_t^{i}|^p] < C_p$, and moreover, $ \displaystyle \sup_{t \in [0,T]} \mathbb{E}[W_p^q(\mu_t^N, \delta_0)] < C$ for $p \le q$.
    
 \item[(ii)] $\mathbb{E}[|X_{t_{j+1}}^{i} - X_{t_j}^{i}|^p] \leq C_p \Delta_n^{p/2}$.
    
 \item[(iii)] $\sup_{t\in [0,T]}\mathbb{E}[W_2^p(\mu_{t}^N, \mu_{t})]\to 0.$ for any $p>0$.
 
\end{enumerate}
\end{lemma}

\subsection{Proof of Theorem \ref{thrm1}}

\noindent
We restrict our attention to the case $d=1$ and set $f:=a_1^4$; the remaining cases follow from polarization identity $xy=2^{-1}((x+y)^2-x^2-y^2)$. We consider the following decomposition
\begin{align}
\widehat{\Lambda}- \Lambda &= \frac{\Delta_n}{N} \sum_{i=1}^N \sum_{j=1}^n f(X_{t_j}^i, \mu_{t_j}^N) - \int_0^T \int_{\R} f(x, \mu_s) \mu_s(dx)  ds \\
& = \frac{1}{N} \sum_{i=1}^N \int_0^T f(X_{s}^i, \mu_{s}) ds - \int_0^T \int_{\R} f(x, \mu_s) \mu_s(dx)  ds \\
&+ \frac{\Delta_n}{N} \sum_{i=1}^N \sum_{j=1}^n f(X_{t_j}^i, \mu_{t_j}) - \frac{1}{N} \sum_{i=1}^N \int_0^T f(X_{s}^i, \mu_{s}) ds
\\
&+ \frac{\Delta_n}{N} \sum_{i=1}^N \sum_{j=1}^n \left[f(X_{t_j}^i, \mu_{t_j}^N) - f(X_{t_j}^i, \mu_{t_j}) \right] \\
& =: M_{\Lambda}^{(1)}+ H_{(1)}+ H_{(2)}
\end{align}
First of all, we note that the term $H_{(1)}$ corresponds to the Riemann sum approximation error associated with stochastic process $f_t:=f(X_t^i,\mu_t)$. According to  \eqref{Itoassa}, 
the stochastic process $a_t:=a(X_t^i,\mu_t)$ is a continuous It\^o semimartingale, and so is $f_t$. Hence, we deduce from  \cite[part 2) of Section 8]{BGJPS} that 
\[
\Delta_n \sum_{j=1}^n f(X_{t_j}^i, \mu_{t_j}) -  \int_0^T f(X_{s}^i, \mu_{s}) ds = \Delta_n A_i^n
\]
with $\E[|A_i^n|]\leq C$ uniformly in $i$ (since $X^i$'s are  identically distributed); in particular, 
this is a consequence of existence of moments of the drift and diffusion coefficients of the process $(a(X_t^i,t))_{t\in[0,T]}$ introduced in \eqref{Itoassa}. Thus, we conclude 
\[
 \frac{1}{N} \sum_{i=1}^N A_i^n = O_{\mathbb{P}}(1).
\]
In other words, $\sqrt{N}H_{(1)}=O_{\mathbb{P}}(\Delta_n \sqrt{N})$ and the latter is negligible as $N\Delta_n^2\to 0$. 

Now, we focus on the term $H_{(2)}$. Here we use the notion of the linear functional derivative and apply it to the function $f$:
\begin{align}
    H_{(2)}=  H_{(2.1)} + H_{(2.2)},
\end{align}
where the terms $H_{(2.1)}$ and $H_{(2.2)}$ are defined via
\begin{align}
    H_{(2.1)}&:= \frac{\Delta_n}{N} \sum_{i=1}^N \sum_{j=1}^n  \int_{\R} \partial_{\mu} f(X^i_{t_j}, y,\mu_{t_j}) (\mu_{t_j}^N - \mu_{t_j})(dy) \\
    H_{(2.2)}&:= \frac{\Delta_n}{N} \sum_{i=1}^N \sum_{j=1}^n  \int_0^1 \int_{\R} 
   \left\{ \partial_{\mu} f(X^i_{t_j}, y,\lambda \mu_{t_j}^N +(1-\lambda) \mu_{t_j}) - \partial_{\mu} f(X^i_{t_j}, y,\mu_{t_j})\right\} \\
   &\times(\mu_{t_j}^N - \mu_{t_j})(dy) d\lambda
\end{align}
For the term $H_{(2.1)}$ we obtain the identity 
\begin{align}
    H_{(2.1)} = \frac{\Delta_n}{N^2} \sum_{i,k=1}^N \sum_{j=1}^n \left( \partial_{\mu} f(X^i_{t_j}, X^k_{t_j},\mu_{t_j}) - \int_{\R} \partial_{\mu} f(X^i_{t_j}, y,\mu_{t_j})   \mu_{t_j}(dy)
     \right).
\end{align}
In the next step, we will use the Hoeffding decomposition for the above $U$-statistics of order $2$. Introduce the function
\[
h_t(y):=\E\left[\partial_{\mu} f(X^1_{t}, y,\mu_{t})\right] -  \int_{\R^2} \partial_{\mu} f(z, y,\mu_{t})  \mu_{t}(dz)  \mu_{t}(dy)
\]
and define the statistic
\[
M_{\Lambda}^{(2)}:=  \frac{\Delta_n}{N} \sum_{i=1}^N \sum_{j=1}^n h_{t_j}(X_{t_j}^i).
\]
Applying standard arguments from Hoeffding decomposition, we conclude that
\begin{align} \label{estima1}
H_{(2.1)}- M_{\Lambda}^{(2)}= O_{\mathbb{P}}(1/N).
\end{align}
At this stage we note that 
\begin{align}
    M_{\Lambda} = M_{\Lambda}^{(1)}+ M_{\Lambda}^{(2)}. 
\end{align}
Due to \eqref{estima1}, we are left to proving that $H_{(2.2)}=o_{\mathbb{P}}(N^{-1/2})$.  To show its asymptotic negligibility,
we first introduce the notation
\[
\mu_t^{N,-k}:= \frac{1}{N-1} \sum_{i=1, i\not = k}^N \delta_{X_t^i}.
\]
For probability measures $\nu_1,\nu_2$, we set
\[
z_{\lambda}(x,y,\nu_1,\nu_2):= \partial_{\mu} 
f(x, y,\lambda \nu_1 +(1-\lambda) \nu_2) - \partial_{\mu} f(x, y,\nu_2).
\]
With this notation in hand, we can decompose the  term $H_{(2.2)}$ as
\[
H_{(2.2)}:=
H_{(2.2.1)} + H_{(2.2.2)},
\]
where the quantities $H_{(2.2.1)}$ and $H_{(2.2.2)}$ are given by
\begin{align}
H_{(2.2.1)}&:= \frac{\Delta_n}{N^2} \sum_{i,k=1}^N 
\sum_{j=1}^n \int_0^{1} \left(
z_{\lambda}\left(X^i_{t_j},X^k_{t_j},\mu_{t_j}^{N,-k},\mu_{t_j}\right) - \int_{\R} 
z_{\lambda}\left(X^i_{t_j},y,\mu_{t_j}^{N,-k},\mu_{t_j}\right) \mu_{t_j}(dy)\right) d\lambda \\
H_{(2.2.2)}&:= \frac{\Delta_n}{N^2} \sum_{i,k=1}^N 
\sum_{j=1}^n \int_0^{1} \left(
z_{\lambda}\left(X^i_{t_j},X^k_{t_j},\mu_{t_j}^{N},\mu_{t_j}\right) - z_{\lambda}\left(X^i_{t_j},X^k_{t_j},\mu_{t_j}^{N,-k},\mu_{t_j}\right)\right) \\
&+ \left(\int_{\R} 
z_{\lambda}\left(X^i_{t_j},y,\mu_{t_j}^{N,-k},\mu_{t_j}\right) -
z_{\lambda}\left(X^i_{t_j},y,\mu_{t_j}^{N},\mu_{t_j}\right) \mu_{t_j}(dy) \right)d\lambda
\end{align}
In the following arguments, we will use the following statement: For any $p>0$ there exists a $C_p>0$ such that
\begin{align} \label{w1stat}
\sup_{\lambda\in [0,1],t\in [0,T], 1\leq k\leq N}
\E\left[W_1\left(\lambda\mu_t^N+(1-\lambda)\mu_t, \lambda\mu_t^{N,-k}+(1-\lambda)\mu_t \right)^p\right]^{1/p}
\leq C_p N^{-1},
\end{align}
where we used that $\E[|X_t^1|^p]<\infty$ for any $t\in [0,T]$
and $p>0$. We also observe the identity $\partial_{\mu}f(x,y,\mu)= 4 a^3(x,\mu) \partial_{\mu}a(x,y,\mu)$. Now, if we use the estimate 
in \eqref{w1stat}, Assumption \ref{assumpt3} and the growth
condition in Assumption \ref{assumpt2}, we readily deduce
the statement
\begin{align}
\E[|H_{(2.2.2)}|]&\leq C \sup_{\lambda\in [0,1],t\in [0,T], 1\leq k\leq N}
\E\left[W_1\left(\lambda\mu_t^N+(1-\lambda)\mu_t, \lambda\mu_t^{N,-k}+(1-\lambda)\mu_t \right)^2\right]^{1/2} \\
&\leq C N^{-1}. \label{H222est}
\end{align}
Now, we focus on the term $H_{(2.2.1)}$. We write
\[
H_{(2.2.1)}=: \frac{1}{N^2} \sum_{i,k=1}^N 
v(i,k)
\]
and we note that $\E[v(i,k)|X^i]=0$ for any $i\not= k$ since the empirical measure $\mu^{N,-k}$ does not contain the $k$th particle. As a consequence, the variance
\[
\text{var}(H_{(2.2.1)})= \frac{1}{N^4} 
\sum_{i_1,i_2,k_1,k_2=1}^N \E[v(i_1,k_1) v(i_2,k_2)]
\]
contains at most $O(N^3)$ non-zero terms as 
$\E[v(i_1,k_1) v(i_2,k_2)]=0$ if the indexes  
$i_1,i_2,k_1,k_2$ are all distinct. Similarly to the estimate
in \eqref{H222est}, using Lemma  \ref{l: momentsss}(iii), we conclude that 
\[
\E[|v(i_1,k_1) v(i_2,k_2)|]\leq C
\sup_{t\in [0,T],1\leq k\leq N}\E[W_1(\mu_t^{N,-k},\mu_t)^4]^{1/2} \to 0.
\]
Summarizing the above, we finally deduce that $H_{(2.2.1)}=o_{\mathbb P}(N^{-1/2})$. 
This completes the proof of Theorem \ref{thrm1}.

\subsection{Proof of Theorem \ref{thrm2}}

We start with the term $\widehat{\Gamma}_k$. We obtain the decomposition
\[
\widehat{\Gamma}_k-\Gamma_k = \Gamma_{k.1} + \Gamma_{k.2}
\]
where the terms $\Gamma_{k.1}$ and $\Gamma_{k.2}$ are defined as
\begin{align}
   \Gamma_{k.1}&:= 
   \frac{\Delta_n}{N} \sum_{i=1}^N \sum_{j=1}^n a_k^2(X_{t_j}^i, \mu_{t_j}^N) a^2(X_{t_j}^i, \mu_{t_j})
   - \int_0^T \int_{\R} a_k^2(x, \mu_t) a^2(x, \mu_t) \mu_t(dx)  dt, \\
   \Gamma_{k.2}&:= \frac{1}{N} \sum_{i=1}^N \sum_{j=1}^n a_k^2(X_{t_j}^i, \mu_{t_j}^N) |X_{t_{j+1}}^i - X_{t_j}^i |^2 - \frac{\Delta_n}{N} \sum_{i=1}^N \sum_{j=1}^n a_k^2(X_{t_j}^i, \mu_{t_j}^N) a^2(X_{t_j}^i, \mu_{t_j}).
\end{align}
Analogously to the proof of Theorem \ref{thrm1} we conclude that
\[
\Gamma_{k.1}= M_k+ o_{\mathbb P}(N^{-1/2}).
\]
Next, we handle the term $\Gamma_{k.2}$. First, we observe 
the decomposition 
\[
X_{t_{j+1}}^i - X_{t_j}^i = a(X_{t_j}^i, \mu_{t_j})
\left(W_{t_{j+1}}^i - W_{t_j}^i \right) +\epsilon_{i,j}
\]
with
\[
\epsilon_{i,j}:= \int_{t_j}^{t_{j+1}} b(X_t^i,\mu_t) dt
+ \int_{t_j}^{t_{j+1}} \left( a(X_t^i,\mu_t) - a(X_{t_j}^i, \mu_{t_j})\right) dW_t^i.
\]
Using this representation and applying the methods of \cite[parts 2) and 3) of Section 8]{BGJPS}, we deduce the decomposition
\begin{align}
    \Gamma_{k.2} = \frac{1}{N} \sum_{i=1}^N \sum_{j=1}^n a_k^2(X_{t_j}^i, \mu_{t_j}^N) 
    a^2(X_{t_j}^i, \mu_{t_j}^N) \left( (W_{t_{j+1}}^i - W_{t_j}^i)^2 - \Delta_n \right) 
    + O_{\mathbb P}(\Delta_n).
\end{align}
By martingale methods, in particular the independence of the increments $W_{t_{j+1}}^i - W_{t_j}^i$ and the $\sigma$-algebra $\mathcal{F}_{t_j}$,
we immediately obtain that
\[
\text{var}(\Gamma_{k.2}) = N^{-1}\Delta_n.
\]
This implies that $\sqrt{N}\Gamma_{k.2} = o_{\mathbb P}(1)$.

The term $\widehat{\mathcal{B}}$ is handled the same way as $\widehat{\Gamma}_k$, which completes the proof of 
Theorem \ref{thrm2}.

\subsection{Proof of Proposition \ref{propSig}}
Recalling the definitions \eqref{defhatv} and \eqref{defv}, we introduce the statistic

\begin{align}
\widehat{\Sigma}^{\prime}
:=
\frac{1}{K}
\sum_{k=1}^K
\frac{N}{K}
\bigl(\widehat{V}_k-V\bigr)
\bigl(\widehat{V}_k-V\bigr)^\top.
\end{align}
Due to Corolloary  \ref{thrm3} and the proof steps in 
Theorems \ref{thrm1} and \ref{thrm2}, we conclude 
the convergence of mixed moments for a fixed $k$:
\begin{align} \label{convofmome}
\E\left[ \left(\sqrt{\frac{N}{K}} \left(\widehat{V}_k-V\right)  \right)^{\otimes p} \right] \to C_p \in \R^{\otimes p}, \qquad p=2,4,
\end{align}
as $N/K\to \infty$, where $C_p=\E[(\mathbf{Z}^*)^{\otimes p}]$. The same statement holds for the statistic 
$\sqrt{N}(\widehat{V}-V)$. This is implied by the error estimates in the proofs of Theorems \ref{thrm1} and \ref{thrm2}, as well as existence of moments of the drift and diffusion coefficients of the process $(a(X_t^i,t))_{t\in[0,T]}$ introduced in \eqref{Itoassa}. As a consequence of 
\eqref{convofmome}, we deduce that 
\begin{align}
    \E[\|\widehat{\Sigma}^{\prime} -\widehat{\Sigma}\|_{\max}]
    &\leq \frac{1}{K}
\sum_{k=1}^K
\frac{N}{K}
\E\left[\left \|\bigl(\widehat{V}_k-V\bigr)
\bigl(\widehat{V}_k-V\bigr)^\top -\bigl(\widehat{V}_k-\widehat{V}\bigr)
\bigl(\widehat{V}_k-\widehat{V}\bigr)^\top\right\|_{\max} \right] \\[1.5 ex]
&\leq \frac{2}{K} \sum_{k=1}^K
\frac{N}{K} 
\E\left[\left \|\widehat{V}_k-V\right\|_{\max}^2 \right]^{1/2}
\E\left[\left \|\widehat{V}-V\right\|_{\max}^2 \right]^{1/2}
\\[1.5 ex]
&= 
\frac{2N}{K} 
\E\left[\left \|\widehat{V}_1-V\right\|_{\max}^2 \right]^{1/2}
\E\left[\left \|\widehat{V}-V\right\|_{\max}^2 \right]^{1/2}
\\[1.5 ex]
& \xrightarrow{\P} 0
\end{align}
as $N/K\to \infty$. 
In the next step, we note that $\widehat{\Sigma}^{\prime}$ is a sum of i.i.d random variables. It holds that 
\[
\E[\widehat{\Sigma}^{\prime}] = 
\frac{N}{K} \E\left[
\bigl(\widehat{V}_1-V\bigr)
\bigl(\widehat{V}_1-V\bigr)^\top \right]
\xrightarrow{\P} \Sigma.
\]
On the other hand, for any $1\leq j,l\leq d^2+d+1$, we deduce that
\[
\text{var}(\widehat{\Sigma}^{\prime}_{jl})
= \frac{N^2}{K^3} \text{var}\left( (\widehat{V}_1-V)_j 
(\widehat{V}_1-V)_l\right) \leq \frac{C}{K} \xrightarrow{\P} 0,
\]
as $K\to\infty$, which again follows from \eqref{convofmome}.

\end{document}